\documentclass[12pt]{article}
\usepackage{graphicx}
\usepackage{subcaption}
\usepackage{caption}
\usepackage{amsmath}
\usepackage{booktabs}
\usepackage{rotating}

\newcommand\pubnumber{CIPANP2015-Chang}
\newcommand\pubdate{\today}

\def\napoli{Dept. of Physics, University of Illinois at Urbana-Champaign, Urbana, IL 61801\\
Lawrence Berkeley National Laboratory, Berkeley, CA 94720}
\def\support{\footnote{Work supported in part by the University Research Association Visiting Scholars' program, the Fermilab Fellowship of Theoretical Physics, and the DOE grant under grant number DOE DE-FG02-13ER42001. Fermilab is operated by Fermi Research Alliance, LLC, under Contract No. DE-AC02-07CH11359 with the U.S. Department of Energy.}}

\def\Title#1{\begin{center} {\Large #1 } \end{center}}
\def\Author#1{\begin{center}{ \sc #1} \end{center}}
\def\Address#1{\begin{center}{ \it #1} \end{center}}

\newcommand\pubblock{\rightline{\begin{tabular}{l} \pubnumber\\
         \pubdate  \end{tabular}}}
\newenvironment{Abstract}{\begin{quotation}  }{\end{quotation}}
\newenvironment{Presented}{\begin{quotation} \begin{center} 
             PRESENTED AT\end{center}\bigskip 
      \begin{center}\begin{large}}{\end{large}\end{center} \end{quotation}}
\def\Acknowledgements{\bigskip  \bigskip \begin{center} \begin{large}
             \bf ACKNOWLEDGEMENTS \end{large}\end{center}}

\def\beq{\begin{equation}}
\def\eeq#1{\label{#1}\end{equation}}
\def\eeqn{\end{equation}}

\def\beqa{\begin{eqnarray}}
\def\eeqa#1{\label{#1}\end{eqnarray}}
\def\eeqan{\end{eqnarray}}

\let\bar=\overbar

\def\Dslash{\not{\hbox{\kern-4pt $D$}}}
\def\dslash{\not{\hbox{\kern-2pt $\del$}}}

\def\msb{{\bar{\ssstyle M \kern -1pt S}}}

\begin{document}
\begin{titlepage}
\pubblock

\vfill
\Title{Hadronic matrix elements of neutral-meson mixing through lattice QCD}
\vfill
\Author{ Chia Cheng Chang\support \\ Fermilab lattice and MILC collaborations}
\Address{\napoli}
\vfill
\begin{Abstract}
Neutral-meson mixing is loop suppressed in the Standard Model, leading to the possibility of enhanced sensitivity to new physics. The uncertainty in Standard Model predictions for $B$-meson oscillation frequencies is dominated by theoretical uncertainties within the short-distance $B$-meson hadronic matrix elements, motivating the need for improved precision. In $D$-meson mixing, the Standard Model short-distance contributions are further suppressed by the GIM mechanism allowing for the possibility of large new physics enhancements. A first-principle determination of the $D$-meson short-distance hadronic matrix elements will allow for model-discrimination between the new physics theories. I review recently published and ongoing lattice calculations of hadronic matrix elements in $B$ and $D$-meson mixing with emphasis on the Fermilab lattice and MILC collaboration effort on the determination of the $B$ and $D$-meson mixing hadronic matrix elements using the methods of lattice QCD.
\end{Abstract}
\vfill
\begin{Presented}
CIPANP2015\\
Vail, Colorado, May 19--24, 2015
\end{Presented}
\vfill
\end{titlepage}
\def\thefootnote{\fnsymbol{footnote}}
\setcounter{footnote}{0}

\section{Motivation}

In the Standard Model (SM), neutral-meson mixing is loop suppressed due to the absence of a tree-level flavor changing neutral current, resulting in greater sensitivity to physics beyond the SM. Historically the neutral Kaon and $B$-meson system had inferred the charm quark and top quark masses respectively before direct production. In the pursuit of indirect detection through precision physics, current and planned experiments from LHCb, BELLE II and BES III promise continual effort in the flavor frontier. Therefore, in tandem with current experimental measurements, and in anticipation of future results, we calculate the hadronic matrix elements that participate in neutral $B$ and $D$-meson mixing using the methods of lattice QCD.

\section{Neutral-meson mixing phenomenology}
\begin{figure}[htb]
\centering
\begin{subfigure}[t]{.45\textwidth}
  \centering
  \includegraphics[width=\linewidth]{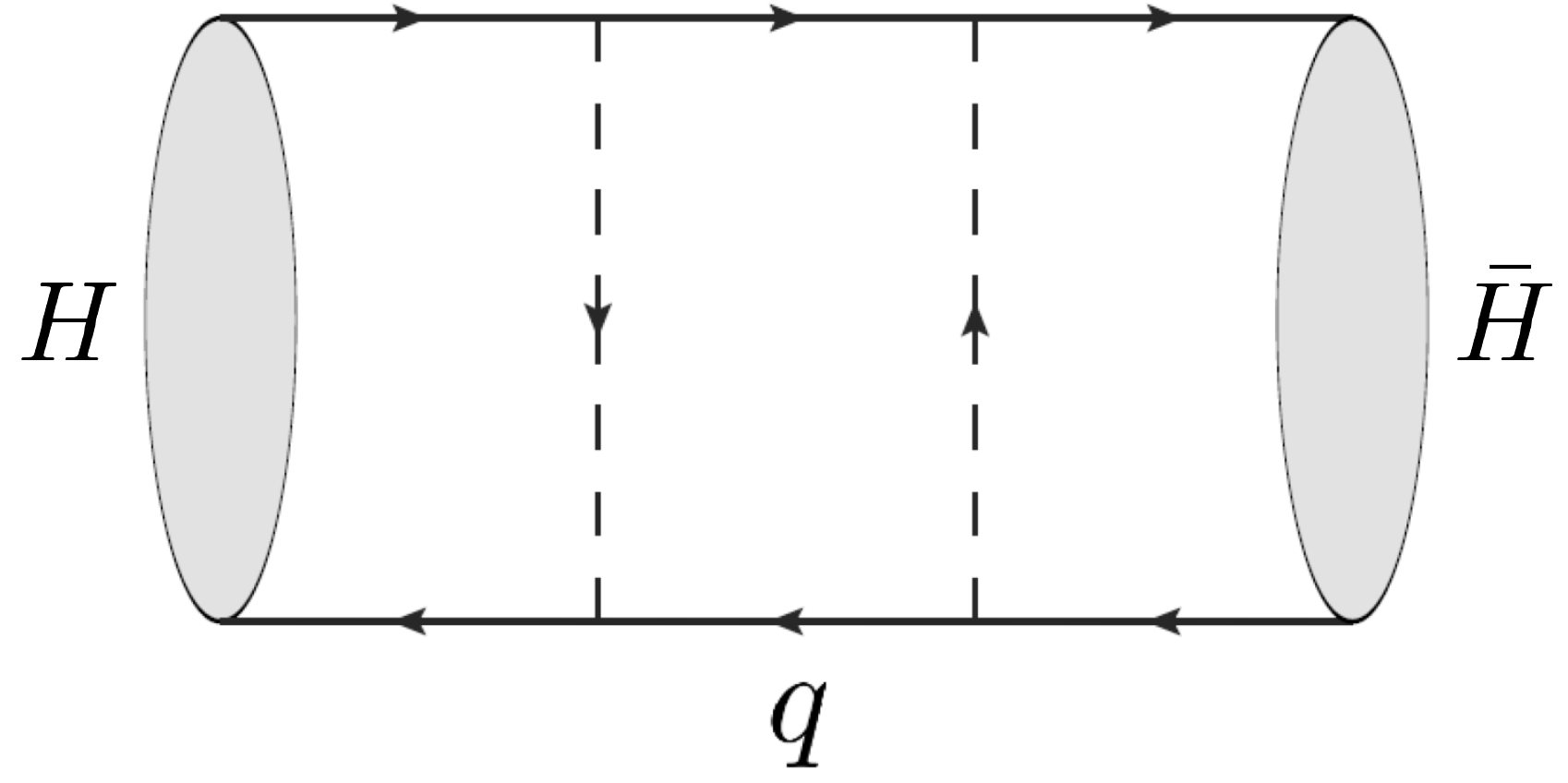}
  \caption{Short-distance diagram. The external state $H$ indicates a general heavy-light neutral-meson. In the SM, the internal quarks $q$ are up-type for $B$- and down-type for $D$-meson mixing.}
  \label{fig:Pheno_Bboxdiagram}
\end{subfigure}\hfill
\begin{subfigure}[t]{.5\textwidth}
  \centering
  \includegraphics[width=\linewidth]{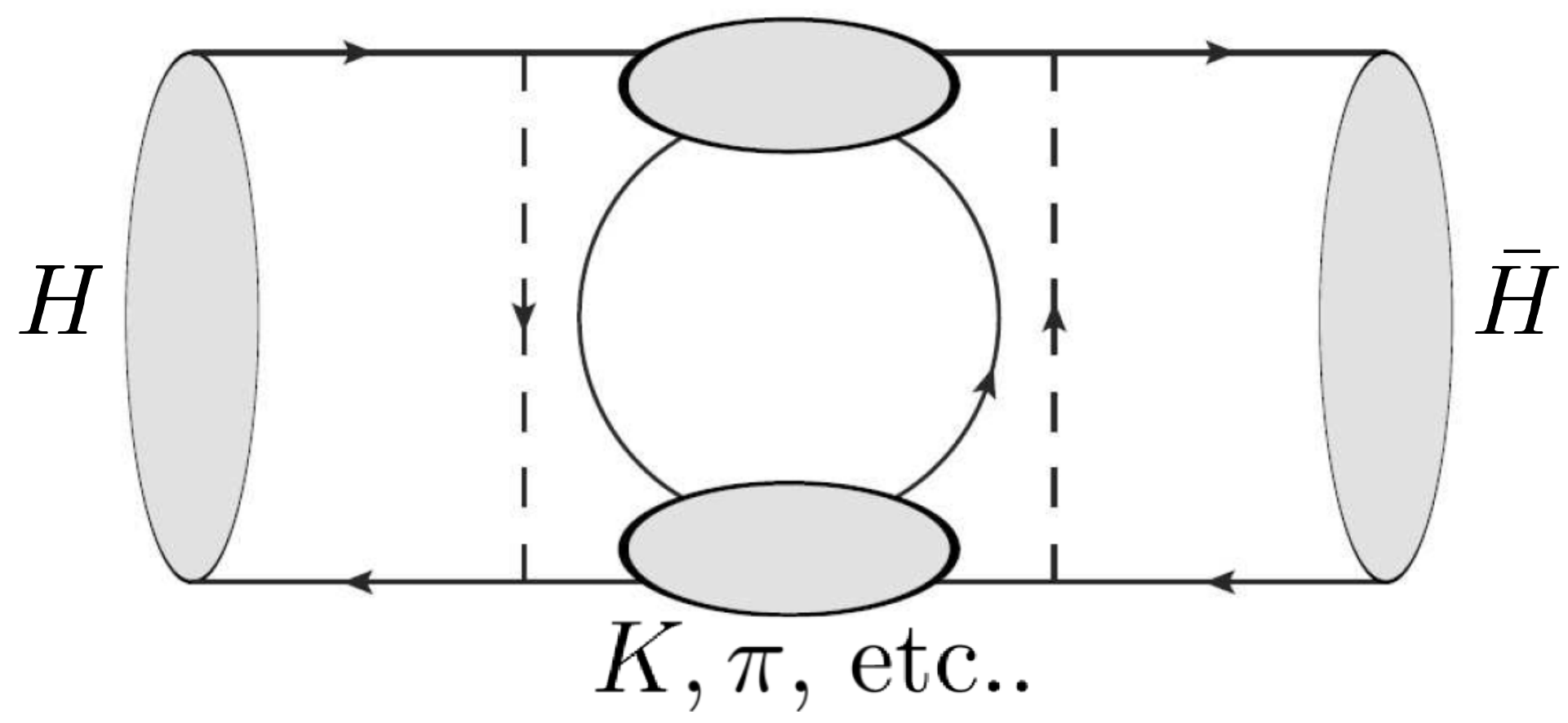}
  \caption{Long-distance diagram. The external state $H$ indicates a general heavy-light neutral-meson. Dominantly a SM process, this diagram proceeds via numerous multi-hadron and resonant intermediate states.}
  \label{fig:Pheno_Longdiagram}
\end{subfigure}
\caption{Contributions to neutral-meson mixing.}
\end{figure}

Mixing of the neutral-meson states in general receives contributions from short- and long-distance diagrams. The Feynman diagram of the short-distance contribution is sketched in Fig.~\ref{fig:Pheno_Bboxdiagram}. The initial and final state hadrons are described by strong dynamics at the MeV scale of $\Lambda_{\text{QCD}}$; in the SM, mixing occurs at the weak scale and is in comparison, a short-distance interaction. For the $B$-meson, the SM box-diagram is enhanced by the top-quark by a factor of $(m_t^2/m_W^2)^2\sim 20$, and is the dominant process describing $B$-meson mixing. Diagrams mediated by the up quark and charm quark are subdominant and are further suppressed by the GIM mechanism. For $D$-meson mixing, the diagram mediated by the bottom quark is CKM suppressed by $\lambda^{10}\sim 0.2^{10}$, while diagrams proceeding via the down quark and strange quark are GIM suppressed. Therefore, $D$-meson mixing in the SM receives subdominant short-distance contributions.

Neutral-meson mixing receives long-distance SM contributions mediated by intermediate hadronic states, as shown in Fig.~\ref{fig:Pheno_Longdiagram}. For $B$-meson mixing, the top-quark can not hadronize, while hadronization via the up quark and charm quark are CKM suppressed by at least $\lambda^4 \sim 0.2^4$, and therefore is a subdominant contribution. On the $D$-meson side, while the bottom quark receives the same CKM suppression as seen in the short-distance diagram, intermediate hadrons with down quarks and strange quarks are only doubly Cabbibo suppressed. Several model dependent estimates have shown that these long-distance contributions may reach the observed level of $D$-meson mixing~\cite{Petrov}. However, the exclusive determination of the long-distance diagram is complicated by many-body multi-channel and resonant intermediate states, while inclusive calculations rely on the poor expansion parameter of $\Lambda_{\text{QCD}}/m_c$. In both cases, estimates are beset by large hadronic uncertainties.

Beyond the SM, neutral-meson mixing proceeds via short-distance diagrams since the scale of new physics is expected to be above the electroweak scale. New physics may facilitate $B$-meson mixing, and can be uncovered by observing a discrepancy between observed $B$-meson mixing and the SM prediction. $D$-meson mixing has ample room to proceed dominantly through physics beyond the SM due to a weakly constrained SM description. Since the phase between the SM and new physics contribution is unknown, one assumes that the SM contribution is subdominant and derive bounds on new physics~\cite{Golowich}. Therefore, the lattice community is motivated to provide determinations of all hadronic matrix elements pertinent to $B$- and $D$-meson mixing at the level of precision comparable to experimental observation.

All hadronic matrix elements involved in neutral-meson mixing can be expressed in a basis of five 4-quark operators,
\begin{align}
\mathcal{O}_1=\bar{\Psi}^\alpha \gamma^\mu L \psi^\alpha \bar{\Psi}^\beta \gamma^\mu L\psi^\beta, &&
\mathcal{O}_2=\bar{\Psi}^\alpha L \psi^\alpha \bar{\Psi}^\beta L \psi^\beta, &&
\mathcal{O}_3=\bar{\Psi}^\alpha L \psi^\beta \bar{\Psi}^\beta L \psi^\alpha,
\label{Pheno_sm}
\end{align}
\begin{align}
\mathcal{O}_4=\bar{\Psi}^\alpha L \psi^\alpha \bar{\Psi}^\beta R \psi^\beta, &&
\mathcal{O}_5=\bar{\Psi}^\alpha L \psi^\beta \bar{\Psi}^\beta R \psi^\alpha,
\label{Pheno_bsm}
\end{align}
where $\Psi$ and $\psi$ denote the heavy and light valence quarks of the $B$ or $D$-meson, $L$ and $R$ are left and right-handed projection operators, and the Greek indices label color degrees of freedom. Operator $\mathcal{O}_1$ has the SM charged current structure, and is directly related to the matrix element of the box diagram. Operators $\mathcal{O}_2$ and $\mathcal{O}_3$ couple only to left-handed quarks, and enter in the calculation of the width difference~\cite{Nierste}. Operators $\mathcal{O}_4$ and $\mathcal{O}_5$ couple to right-handed quarks and are in general needed for new physics models. Therefore, on the lattice we need only calculate five matrix elements for each neutral-meson to describe all hadronic interactions.

\section{Lattice calculation}

\begin{figure}[htb]
\centering
\begin{subfigure}[t]{.49\textwidth}
  \centering
  \includegraphics[width=\linewidth]{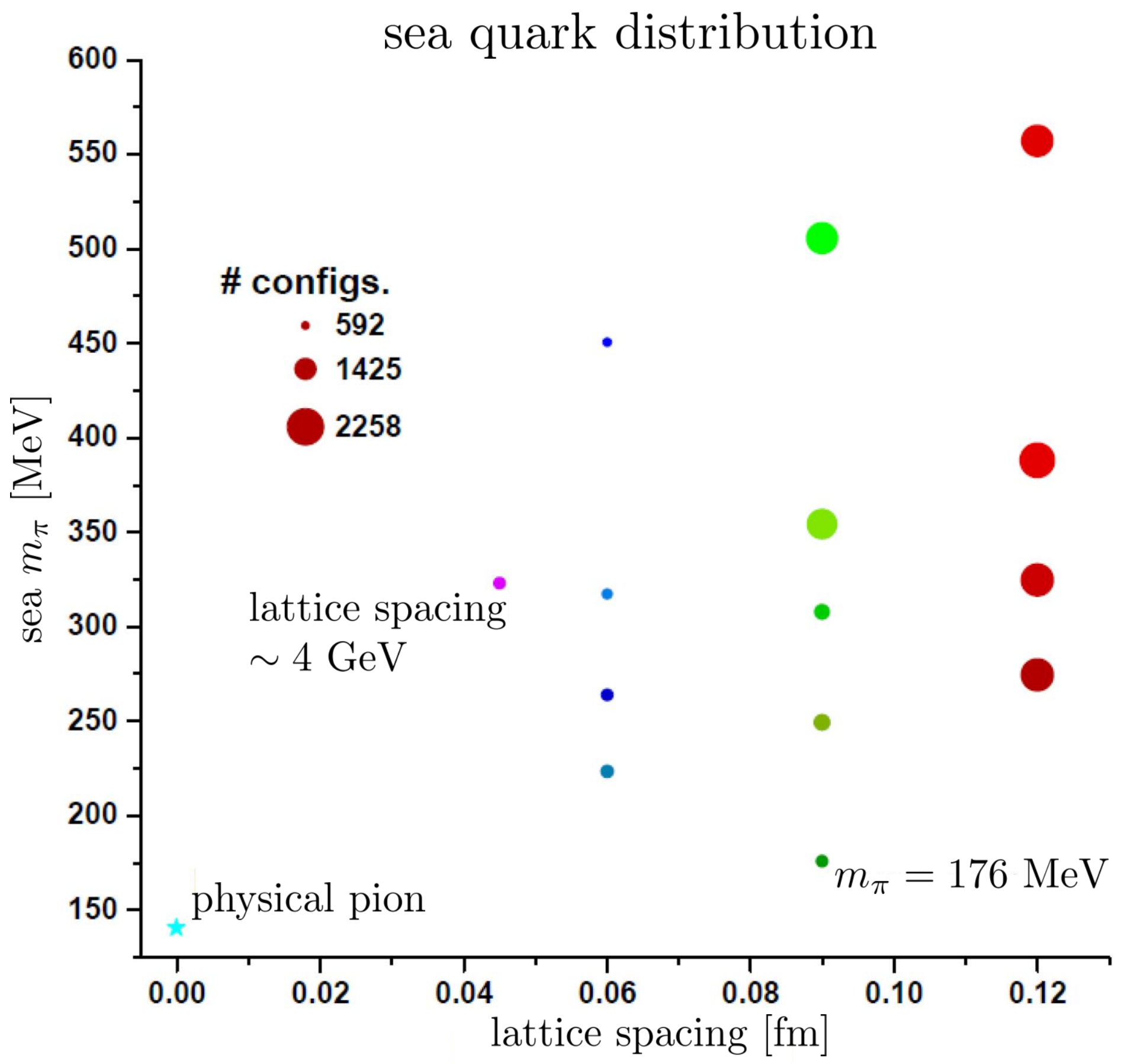}
  \caption{MILC asqtad gauge ensembles. Colors label ensembles with different lattice spacing, and darker hues correspond to lighter pion masses. The cyan star marks the physical pion mass in the continuum limit. Ensembles include 592 to 2258 uncorrelated Monte Carlo samples indicated by the size of the markers.}
  \label{fig:MILCgauge}
\end{subfigure}\hfill
\begin{subfigure}[t]{.49\textwidth}
  \centering
  \includegraphics[width=\linewidth]{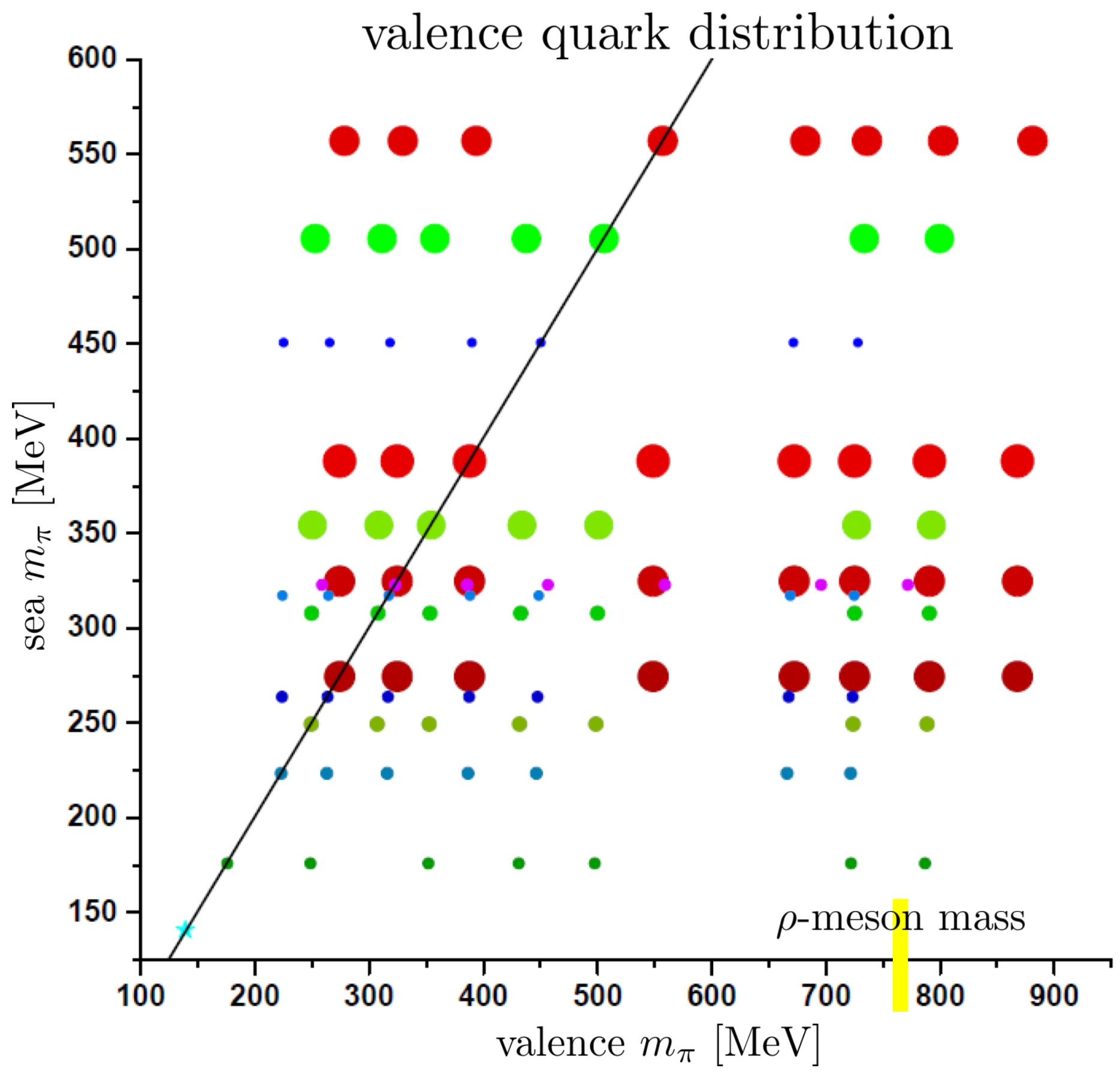}
  \caption{Light-quark propagators. Statistically correlated light-quark propagators of different valence masses are obtained at each gauge configuration. The black diagonal line passes through $m_{\text{sea}} = m_{\text{valence}}$. A total of 102 data points are available per matrix element to guide the chiral-continuum extrapolation.}
  \label{fig:Valence}
\end{subfigure}
\caption{Dataset used for the analysis.}
\label{fig:dataset}
\end{figure}

\begin{figure}[htb]
	\centering
		\includegraphics[width=0.70\textwidth]{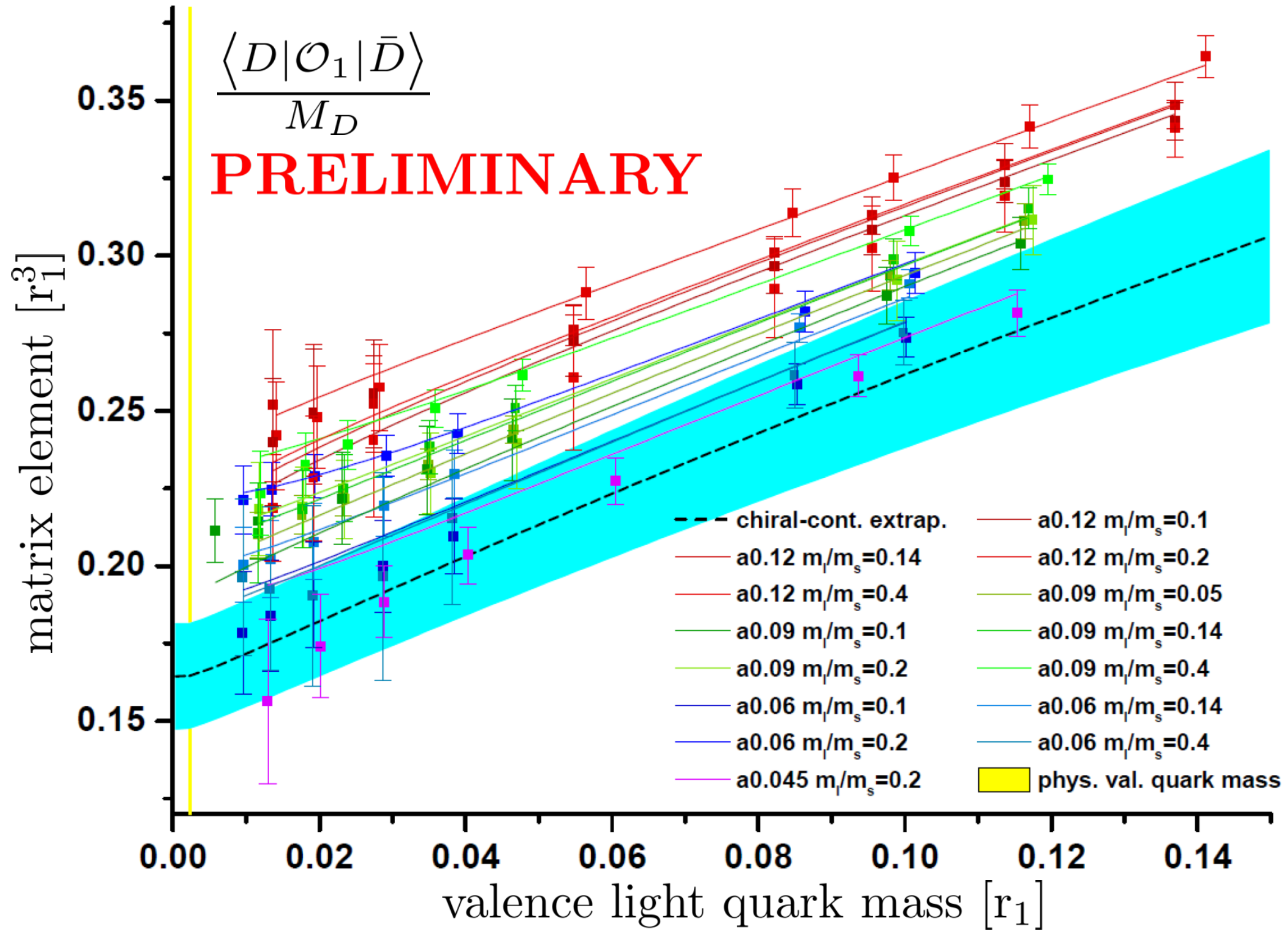}
	\caption{Chiral-continuum extrapolation. Extrapolation to the physical point for the $D$-meson mixing matrix element of operator $\mathcal{O}_1$. The data points share the color coding of Fig.~\ref{fig:dataset} to reflect the gauge ensemble in which they are evaluated on. The continuum extrapolation is marked by the cyan band, the yellow line marks the physical up-quark mass.}
	\label{fig:chiralfit}
\end{figure}

The Fermilab lattice collaboration calculates hadronic matrix elements for $B_d$, $B_s$, and $D$ using a large set of the MILC collaboration's asqtad gauge ensembles~\cite{MILC}. The light valence quark propagators are generated using the asqtad action, and the heavy-valence quark propagators use the Sheikoleslami-Wohlert action with the Fermilab interpretation~\cite{Aida}. A detailed list of ensembles used is tabulated in Ref.~\cite{Lat13}. Illustrated in Fig.~\ref{fig:MILCgauge}, the wealth of data over a wide range of quark masses and lattice spacings allows us to control and estimate all systematic errors. A summary of the inclusion of various systematic errors is detailed in Ref.~\cite{Lat14D,Lat14B}. We perform a combined chiral-continuum extrapolation over the entire data set and obtain the matrix elements at the physical point~\cite{Claude}, as illustrated in Fig.~\ref{fig:chiralfit}. Preliminary results reveal that the Fermilab lattice calculation for matrix elements have errors ranging from 5\% to 15\%; a detailed error breakdown is arranged in Tab.~\ref{tab:Errbudget}.

\begin{table}[htb]
\footnotesize
\begin{center}
\begin{tabular}{|c|ccccc|}
  \toprule
    & $\left<H|\mathcal{O}_1|\bar{H}\right>$ & $\left<H|\mathcal{O}_2|\bar{H}\right>$ & $\left<H|\mathcal{O}_3|\bar{H}\right>$ & $\left<H|\mathcal{O}_4|\bar{H}\right>$ & $\left<H|\mathcal{O}_5|\bar{H}\right>$\\
   \midrule
   $B_d$ & 7.7\%& 8.4\%& 15.2\%& 6.7\%& 8.9\%\\

   $B_s$ & 5.5\%& 5.8\%& 10.8\%& 5.1\%& 6.6\%\\

   $D$ & 8.7\%& 4.7\%& 6.1\%& 5.0\%& 9.1\%\\
   \bottomrule
\end{tabular}
\caption{Preliminary relative error for the $B_d$, $B_s$, and $D$-meson hadronic matrix elements.}
\label{tab:Errbudget}
\end{center}
\end{table}

\clearpage
\section{Phenomenological implications}
\subsection{$B$-meson mass splitting}
In the SM, the $B$-meson mass splitting is given by,
\begin{equation}
\Delta M_q = \left(\frac{G_F^2 M_W^2 S_0}{4\pi^2 M_{B_q}}\right)\eta_B(\mu)\left<B_q|\mathcal{O}_1|\bar{B}_q\right>(\mu)
\end{equation}
where the factor in the parenthesis and $\eta_B$ are know electroweak contributions with QCD corrections. In combination with $\left<B_q|\mathcal{O}_1|\bar{B}_q\right>$, the theoretical prediction can then be compared with the experimental observation of $\Delta M_q$.

The matrix elements are often expressed in terms of the bag parameter,
\begin{equation}
\left<H|\mathcal{O}_i|\bar{H}\right> = \mathcal{C}_i M_H^2f_H^2\hat{B}^{(i)}_H,
\end{equation}
where $i$ labels one of five 4-quark operators, $M_H$ is the mass of the heavy-light meson $H$, $f_H$ is the meson decay constant, $\hat{B}^{(i)}_H$ is the bag parameter, and $\mathcal{C}_i$ is a factor that normalizes $\hat{B}^{(i)}_H$ in the vacuum saturation approximation.

The latest experimental observations of the $B$-meson mass differences are
\begin{align}
\Delta M_d = 0.510(3) \text{ ps}^{-1} && \Delta M_s = 17.761(22) \text{ ps}^{-1},
\end{align}
with respective errors of 0.6\% and 0.1\%~\cite{PDG}. The current FLAG-averaged lattice determination of the bag parameters $\hat{B}_{B_{d,s}}$ comes entirely from HPQCD with 8\% and 5\% errors~\cite{HPQCDB,FLAG}. The Fermilab lattice collaboration will provide an independent determination of the bag parameters, however preliminary results suggest similar errors. There are ongoing efforts to reduce the hadronic uncertainty with hints of sub-percent precision for ratio quantities~\cite{HPQCDprelim}.

\subsection{Unitarity triangle constraint}
\begin{figure}[htb]
	\centering
		\includegraphics[width=0.70\textwidth]{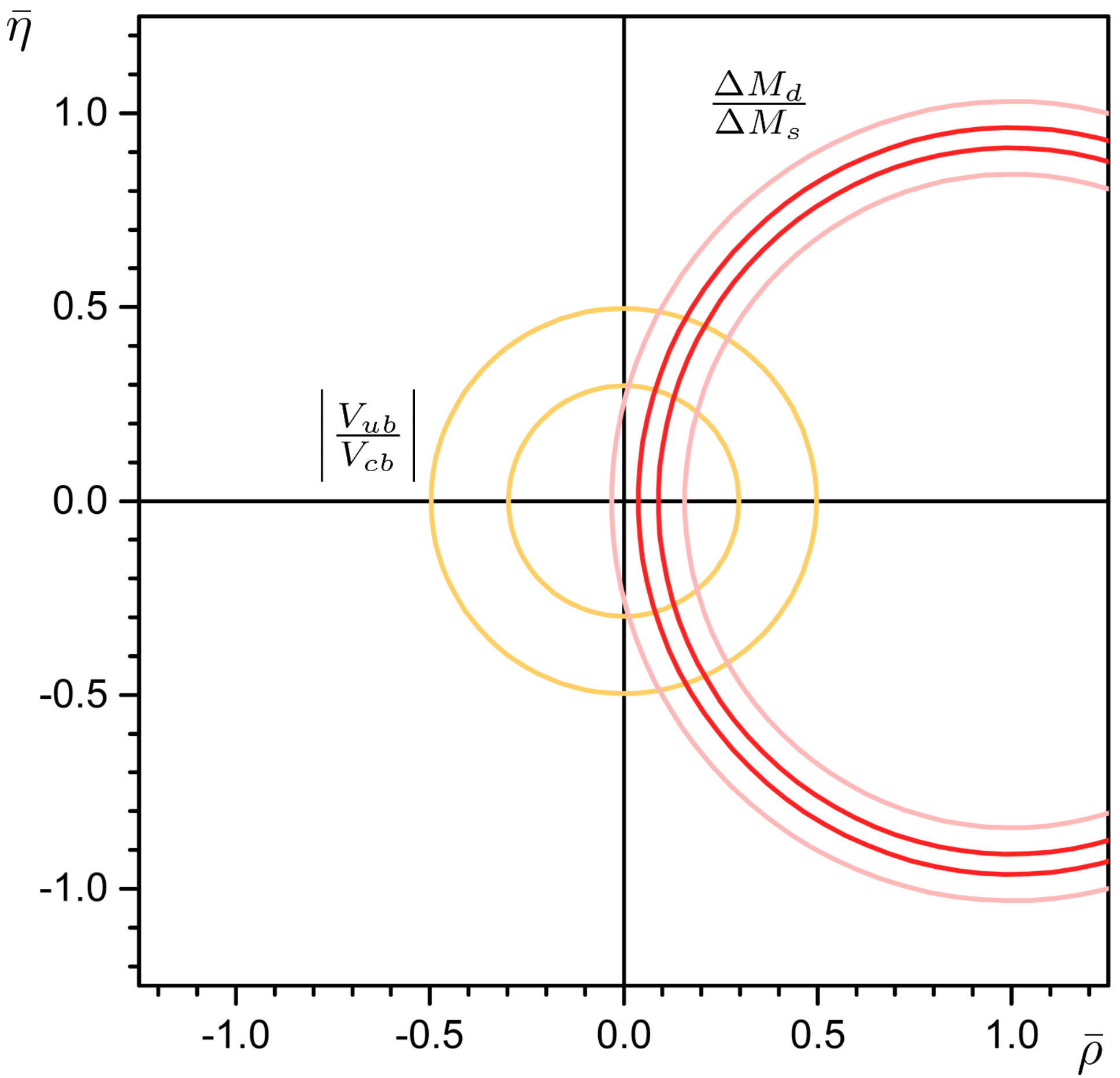}
	\caption{Unitarity triangle constraints from semileptonic $B$ decays and $B$-meson mixing at two standard deviations. Pink curves are evaluated at the FLAG average $\xi$ value, red curves are evaluated at the FLAG average central value with the Fermilab lattice preliminary error. All other parameters are taken from Ref.~\cite{PDG}.}
	\label{fig:UT_annotated}
\end{figure}

The $B$-meson mass splittings constrain one side of the unitarity triangle through the following relationship
\begin{equation}
\left|\frac{V_{td}}{V_{ts}}\right|=\xi\sqrt{\frac{\Delta M_d M_{B_s}}{\Delta M_s M_{B_d}}},
\end{equation}
where $\xi^2$ is the ratio of the $B_s$ and $B_d$ matrix elements for operator $\mathcal{O}_1$. On the lattice, $\xi$ is better determined than the bag parameters since taking the ratio partially cancels statistical and systematic errors. The current FLAG average of $\xi$ comes entirely from Fermilab lattice with 5\% error~\cite{FLAG,XiPaper}. Preliminary results from Fermilab lattice reduce this error down to 1\%, largely as a result of enlarging the data set and controlling the systematic error stemming from wrong-spin taste-mixing effects~\cite{Lat12}. The projected improvement to the constraint on the unitarity triangle is illustrated in Fig.~\ref{fig:UT_annotated}, where the pink bounds shrink to the red bounds.

\subsection{$D$-meson mass splitting}
Assuming that $D$-meson mixing proceeds dominantly through BSM processes, the mass splitting is given by,
\begin{equation}
\Delta M = \mathcal{C}_{\text{NP}}^i\left<D\left|\mathcal{O}_i\right|\bar{D}\right>
\end{equation}
where $\mathcal{C}_{\text{NP}}^i$ is the model-dependent short-distance coefficient. Supplied with experimental observation for $\Delta M$, and lattice calculations for $\left<\mathcal{O}_i\right>$, parameters in new physics models can be constrained. One may also provide model independent constraints to new physics following Ref.~\cite{Bona} and~\cite{ETMC}.

\begin{figure}[htb]
	\centering
		\includegraphics[width=1.00\textwidth]{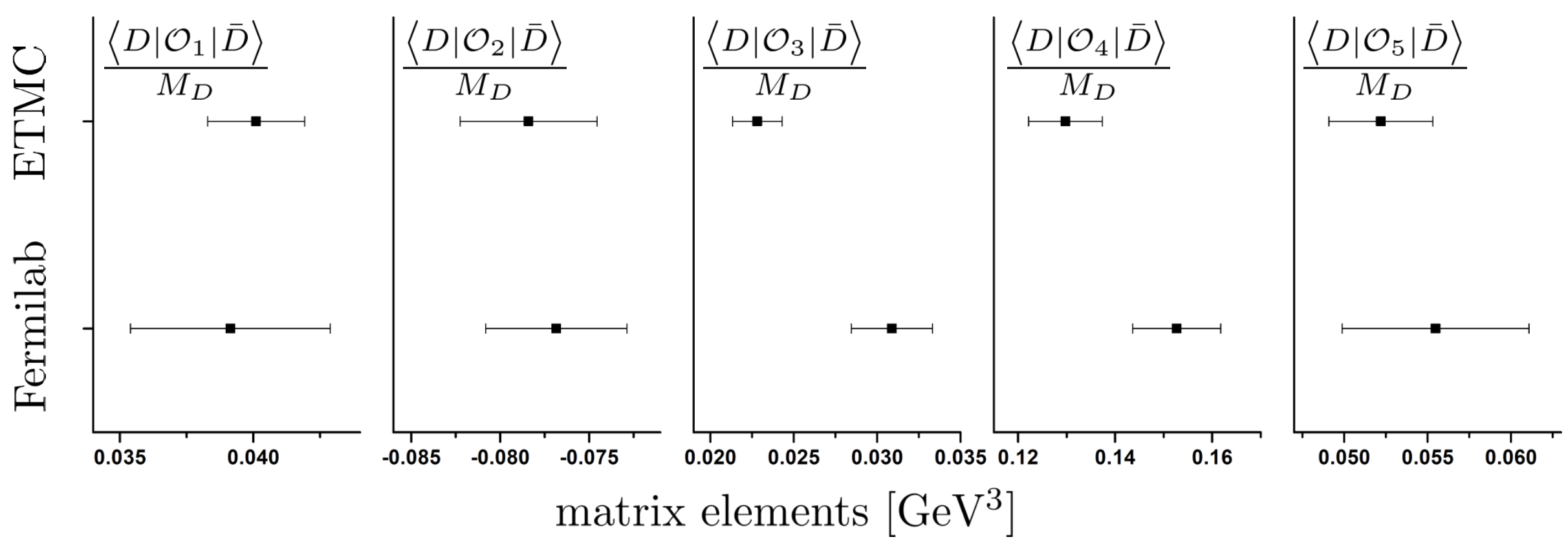}
	\caption{$D$-meson mixing matrix elements from ETMC and preliminary Fermilab lattice results. Bag parameters from ETMC~\cite{ETMC} are converted to matrix elements with the $D$-meson decay constant from Ref.~\cite{ETMCfD}.}
	\label{fig:comparison}
\end{figure}

The current experimental measurement for the $D$-meson mass splitting is
\begin{equation}
\Delta M = 0.0044(20)\text{ ps}^{-1}
\end{equation}
with a 45\% error~\cite{PDG}. Currently there is one published result on the $D$-meson mixing hadronic matrix elements from the ETMC with 2 dynamical sea quarks~\cite{ETMC}. Results from Fermilab lattice will be the first 2+1 dynamical sea quark calculation and preliminary results, shown in Fig.~\ref{fig:comparison}, suggest the calculations are statistically consistent.

Planned experiments for the next decade indicate that the error is expected to be reduced to the 10\% level, on par with lattice uncertainties of Tab.~\ref{tab:Errbudget}.

\section{Conclusions and outlook}
We are calculating $B_d$, $B_s$, and $D$-meson hadronic matrix elements with 5\% to 10\% relative error, and the ratio quantity $\xi$ at the 1\% level. For the $B$ system, hadronic quantities are currently the dominant source of uncertainty and ongoing efforts from the lattice community aim to obtain errors commensurate with experiment. Uncertainty on the $D$-meson hadronic matrix elements are comparable to future experimental observation over the next decade. However, uncontrolled hadronic uncertainty exists in the SM long-distance contribution.

\Acknowledgements
I would like to thank the organizers and conveners for inviting me to speak at this wonderful conference, and to my collaborators: Claude Bernard, Chris Bouchard, Aida El-Khadra, Elizabeth Freeland, Elvira G\'amiz, Andreas Kronfeld, Jack Laiho, and Ruth Van de Water, as well as the rest of the members of the Fermilab lattice and MILC collaborations.

\end{document}